\documentclass[vecphys]{svmult}

\usepackage{makeidx}         
\usepackage{graphicx}        
\usepackage{multicol}        
\usepackage[bottom]{footmisc}

\usepackage{amsmath}
\usepackage{amssymb,bm}

\DeclareMathOperator{\Tr}{Tr}
\newcommand\dd{{\mathrm d}}
\newcommand\ee{{\mathrm e}}
\newcommand\ii{{\mathrm i}}

\makeindex             

\begin{document}

\title*{Low temperature decoherence and relaxation
in charge Josephson junction qubits}
\titlerunning{Decoherence and relaxation in charge qubits}
\author{Alex Grishin \and Igor V. Yurkevich \and
Igor V. Lerner }

\authorrunning{A. Grishin \and I. V. Yurkevich \and
I. V. Lerner} \institute{School of Physics and Astronomy,
University of Birmingham (UK) }
%
%
\maketitle

\section{Introduction}

Research interest in controllable two level systems, which have
been enthusiastically called quantum bits or qubits, has grown
enormously during the last decade. Behind a huge burst of activity
in this field stands an idea of what is possible in principle but
extremely difficult to achieve instrumentally - the fascinating
idea of quantum computing. The very principle of quantum
superposition allows many operations to be performed on a quantum
computer in parallel, while an ordinary `classical' computer,
however fast, can only handle one operation at a time. The
enthusiasm is not held back by the fact that exploiting quantum
parallelism is by no means straightforward, and there exist only a
few algorithms (e.g., \cite{shor:97,grover}) for which the quantum
computer (if ever built) would offer an essential improvement in
comparison with its `classical' counterpart. Even if other uses of
quantum computing prove limited (which might or might not be the
case), its existence would most certainly lead to a breakthrough
in simulations of real physical many-particle systems.

Whether or not the ultimate goal of building a working quantum
computer is ever achieved, both experimental and theoretical
studies of properties of single or entangled qubits are
flourishing. One of the most compelling reasons for this is an
exciting overlap of the possibility of a future technological
breakthrough and the reality of research in fundamentals of
quantum mechanics. There exist various experimental realisations
of the qubit, amongst which solid state qubits are of particular
interest as they provide one of the most promising routes to
implementing a scalable set of qubits, which is one of the minimal
requirements for implementing quantum computations.  However, any
solid state qubit, albeit representing effectively a two-level
system, comprises a huge number of internal degrees of freedom
whose unavoidable coupling to the environment leads to loss of
coherence. Quantum computations require a set of fully or at least
partially entangled states on which some unitary operations are
performed. Decoherence would make evolution of states non-unitary
and would lead to the loss of entanglement between the states.
Thus, the loss of coherence before a sufficient amount of quantum
operations was performed would be the major impediment in using
solid-state qubits in quantum computations. It is believed that
tens of thousands of unitary operations are required for quantum
computation to become a reality \cite{loss3} so that sufficiently
long  decoherence times (much longer than those currently archived
in the best solid state qubits) should be achieved experimentally.
It necessitates a better theoretical understanding of realistic
mechanisms of decoherence.

In these lectures, after illustrating in a simple way the main
features of decoherence in a generic qubit coupled to the
environment (Section~\ref{sec:dec}),  we will consider onset of
decoherence in one particular realisation of qubit, namely charge
Josephson junction (JJ) qubit. First, we will describe briefly
what is the JJ qubit (Section~\ref{sec:jjqubit}) and then focus on
a mechanism widely believed to be responsible for one of the main
channels of decoherence for the charge JJ qubit, namely its
inevitable coupling to fluctuating background charges
(Section~\ref{sec:fbcmodel}). This model has been thoroughly
investigated in all regimes [4--7]
and has a tutorial
advantage of being exactly solvable within a fully quantum
approach \cite{gyl2} and showing a rich variety of different
regimes with a non-trivial  dependence on temperature and on
(unfortunately ill known experimentally) the strength of coupling
between the qubit and the fluctuating charges.

In Section \ref{sec:main} we will offer a solution for decoherence
rate in this model which is formally exact at an arbitrary
temperature $T$.  In the `high-$T$ regime' (which could still
correspond to rather low temperatures),  the decoherence rate
saturates and becomes $T$-independent, while at low temperature it
turns out to be linear in $T$ and behave non-monotonically as a
function of the coupling strength between qubit and the
environment. In conclusion we will also consider the relaxation
rate, albeit only perturbatively  with respect to the coupling
strength,  and demonstrate that the model can qualitatively
explain the experimentally observed \cite{astafiev} quasi-linear
behaviour of the spectral density of noise with humps at certain
frequencies.

\section{Coupling to the environment and decoherence}
\label{sec:dec}%
Before considering a realistic solid-state qubit, we start with
illustrating what is the loss of coherence in a generic qubit.
Such a qubit is a two-level system so that its Hamiltonian can be
mapped to that of spin $\tfrac12$ and written
\begin{align}\label{H0}
\hat H_0=\tfrac12 B_z\hat \sigma_z-\tfrac12 B_x\hat \sigma_x \,.
\end{align}
where $\bf B$ is an effective `magnetic field' (measured here in
energy units). States of the qubit can be described in terms of
its density matrix,
\begin{align}\label{DM}
{\hat\rho}(t)= \sum_{i,j=\uparrow,\downarrow}^{ }
\left|i\right>\rho_{ij}(t)\left<j\right|\,.
\end{align}
When the qubit (or any system) is in a pure state, one can always
find a basis where $\hat{\rho} =\sum_{i}^{ }\left|i\right>
\left<i\right| $. In any other (rotated) basis, the density matrix
of a pure state obeys $\hat{\rho}^2=\hat{\rho}.$  In a rotated
basis, the density matrix performs a unitary evolution described
by the Heisenberg equation of motion,
\begin{align}\label{HEM}
    \frac{\partial\hat{\rho }(t)}{\partial
t}=-\text{i}\left[\hat{H_0},\hat{\rho}(t)\right]\,,
\end{align}
whose solution is $\hat{\rho}(t)=\hat U\hat{\rho}(0)\hat
U^\dagger$ where in this trivial case the evolution operator
$\hat{U}={\rm e}^{-{\rm i}H_0t}$.
\begin{figure}
\vspace*{2mm}

 \centering
\includegraphics[height=5cm,clip=true]{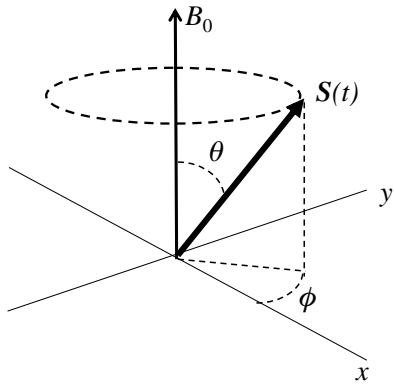}
\caption{Bloch representation} \label{fig:bloch}
\end{figure}
In a semiclassical language, such an evolution is convenient to
visualise using the Bloch sphere representation,  shown in
Fig.\ref{fig:bloch}.  There the spin evolution is parameterised by
two angles on a unit sphere, $\theta$ and $\phi$. For the simple
system described by Eqs.~(\ref{H0}) -- (\ref{HEM}) above, such an
evolution is simply rotating around $z$ axis, provided that the
`field' ${\mathbf B}=\left({B_z, 0}\right)$, and the initial state
of the spin was diagonal in a different basis. The Bloch angles
are related to the density matrix by
\begin{align}\label{rhoS}
     \rho(t)=\frac12
\left({{1+{\mathbf S}(t)\cdot\bm \sigma}}\right) \,.
\end{align}

An inevitable coupling to the environment (bath) can be
schematically described by the Hamiltonian
\begin{align}\label{coupl}
     \hat H= \hat H_0 + \hat V_{\text{coupl}}+\hat H_{\text{bath}}\,.
\end{align}
Now it is the full density matrix of the qubit $+$ bath which
obeys the Heisenberg equation of motion. As we are interested in
states measured on the qubit, we need to introduce a so called
`reduced' density matrix, $\hat{\rho}^q=\Tr_{{\text{bath}}}
\hat{\rho}$ where the trace is taken over all the bath states with
the proper Gibbs weight. We will describe a consistent way of
performing such a trace in Section \ref{sec:main}. Here we want
just to illustrate how the coupling to the environment makes the
time evolution non-unitary and leads to the loss of coherence.

Let us consider a model in which there is a minimal, so called
longitudinal coupling of the qubit to the bath,
\begin{align}\label{Xc}
    V_{{\text{coupl}}}= \hat X\hat{\sigma}_z\,.
\end{align}
  Let us further assume
that the qubit was prepared in a pure state but in a different
basis, so that its density matrix has both diagonal and
off-diagonal elements,
\begin{align*}
    \hat \rho\equiv\begin{pmatrix}
      \rho_{11} & \rho_{12} \\
      \rho_{21} & \rho_{22} \\
    \end{pmatrix}=\begin{pmatrix}
      n & f \\
      f^* & 1-n \\
    \end{pmatrix}\,,
\end{align*}
with $|f|^2=n-n^2$ at $t=0$ assuming that the qubit was prepared
in a pure state, $\hat{\rho}^2=\hat{\rho}$.
 Let finally switch off the $x$ component
of the `magnetic field' $\bf B$. In the absence of coupling, the
spin that represents our qubit on the Bloch sphere of
Fig.~\ref{fig:bloch} would simply rotate at a constant frequency
$\omega_0\propto B_z$ around $z$ axis. The longitudinal coupling
effectively introduces and additional, fluctuating time dependent
component of the field, $B_z(t)$. This fluctuating field results
from the thermal noise in the bath and inevitably destroys
coherence as we will now show.

The spin rotation is described in a semiclassical language by the
Landau-Lifshitz equation,
\begin{align*}
    \frac{\dd {\mathbf S}}{\dd t}=\bf S\times B\,.
\end{align*}
In the absence of the transverse component of the magnetic field,\
the diagonal elements of the density matrix, related to $\bf S$ by
Eq.~(\ref{rhoS}), remain constant. In the presence of the
environment-induced time-dependent component of the longitudinal
magnetic field,  $\bf S$ rotates with a changing frequency
resulting in the off-diagonal elements of $\hat{\rho} $ acquiring
the following time-dependence:
\begin{align}\label{t-dep}
    \rho_{12}(t)=\tfrac{1}{2}\left({S_x+\ii S_y}\right)&= \rho_{12}(0)
 \ee^{\ii\omega_0t +\ii\int_0^tB(\tau)\dd\tau}
\end{align}

The contributions due to the fluctuating field, $B_z(t)$, should
be averaged over time. It is natural to assume that fluctuations
of $B_z$ are Gaussian, as they are due to an very large number of
degrees of freedom in the bath. For Gaussian fluctuations, one can
use the standard averaging formula,
\begin{align*}
      \left<{{\ee^{i\varphi}}}\right>=\ee^{-\frac12
      \left<{{\varphi^2}}\right>}\,,
\end{align*}
so that
\begin{align}\label{rho-av}
   \left<{ {\rho_{12}(t)}}\right>
&=\rho_{12}(0)
 \ee^{\ii\omega_0t } \left<\ee^{
 \ii\int_0^t\dd\tau
  B(\tau) } \right>\notag\\
   &= \notag\rho_{12}(0)
 \ee^{\ii\omega_0t }  \exp{\left[ -\frac12\int_0^t\dd\tau
 \int_0^t\dd\tau'\Big <{ B(\tau) B(\tau')}\Big>\right]}\\
  &= \rho_{12}(0)
 \ee^{i\omega_0t } \exp{\left[ - t\int_{-t}^t\dd\tau  \Big
 <{B(0) B(\tau)}\Big>\right]}\equiv \rho_{12}(0)
 \ee^{i\omega_0t }\ee^{-\Gamma_2t}
\end{align}
Here we have semi-formally defined the decoherence rate $\Gamma_2$
as the rate of relaxation of the off-diagonal part of the density
matrix, using that the correlation function of the fluctuating
field $B_z$ depends only on the time difference. In this
definition, $\Gamma_2$ can still be time-dependent. However, for
large enough $t$, longer than the longest relaxation time for
thermal noise in the bath which makes $B_z$ fluctuating,
$\Gamma_2$ should saturate at some limiting value. Thus, making a
bit more formal definition, we arrive at
\begin{align}\label{G-formal}
    \Gamma_2\equiv\frac{1}{T_2} = \lim_{t\to\infty}
    \int_{-t}^t\dd\tau  \Big
 <{B(0) B(\tau)}\Big>=S_B\left({\omega\approx 0}\right)\,,
\end{align}
where $S_B(\omega)$ is the noise power spectrum of the fluctuating
field $B$,
\begin{align}\label{noise}
    S_B(\omega)\equiv
    \int_{-\infty}^\infty   \Big
 <{B(0) B(t)}\Big>\,\ee^{-\ii \omega t}\dd t\,.
\end{align}
The above calculation should be considered just as an illustration
as we have just introduced Gaussian fluctuations of the `magnetic
field' $B_z$. However, it is very easy both to make this sort of
calculations fully quantum-mechanical, and to generalise the model
beyond the longitudinal coupling.

To show that the effective coupling to the bath leads to the
thermal noise that directly results in the appearance of
decoherence given by Eq.~(\ref{G-formal}), one needs to model the
bath. The most standard theoretical approach to such modeling and
thus to decoherence by the environment is based on spin-boson
models \cite{leggett,breuer}, where the environment is modelled as
a set of harmonic oscillators,
\begin{align}\label{S-B}
    \hat H_{\text{bath}}=\sum_{\bf k}^{ }\omega_{\bf
    k}^{\phantom{\dagger}}
    b^\dagger_{\bf k}  b^{\phantom{\dagger}}_{\bf k}
\end{align}
The linear coupling in Eq.~(\ref{Xc} ) should be understood as
coupling to all the oscillator degrees of freedom,
\begin{align}\label{X-c}
     V_{{\text{coupl}}}=\hat{X}\hat{\sigma}_z\equiv
      \sum_{\bf k}^{ }\left({\lambda_{\bf
    k}^{\phantom{\dagger}}
    b^\dagger_{\bf k}+{\text{h.c.}}}\right) \hat{\sigma}_z
\end{align}
The fluctuation-dissipation theorem allows one to express the
noise power for the coupling operator $\hat{X} (t)$ (in the
interaction picture) as
\begin{align}\label{Xnoise}
 S_X(\omega)=   \left<{X_\omega^2}\right>+\left<{X_{-\omega}^2}\right>
 =2J(\omega)\coth\frac{\hbar \omega}{2T},
\end{align}
where $J(\omega)$ depends on the density of oscillator states,
i.e.\ on the spectrum $\omega_{\bf k}$ in Eq.~(\ref{S-B}), and on
the coupling $\lambda\left({\omega}\right)$ in Eq.~(\ref{X-c}).
The decoherence rate is still given by Eq.~(\ref{G-formal}), where
$S_X(\omega)$ should be substituted for  $S_B(\omega)$. For the
most typical, Ohmic model, $J\left({\omega}\right)\propto \omega$
so that $\Gamma_2\propto T$. As we will show later, such a linear
dependence on temperature is characteristic for a realistic model
to be considered later in these lectures, but only for
sufficiently low $T.$

It is straightforward to generalise our considerations beyond the
longitudinal model. To this end one should take into account the
$\sigma_x$-proportional contribution in the qubit action -- this
would be general enough even if one leaves the coupling as it
stands in Eqs.~(\ref{Xc}) or (\ref{X-c}).

\section{Charge Josephson junction qubit}
\label{sec:jjqubit} Not any two-level system could serve as a
qubit. There is a set of requirements that such a system must
satisfy in the first place. In 1997 David DiVincenzo
\cite{divincenzo2} formulated five criteria which have to be
satisfied by a physical system considered as a candidate for
quantum computation, conditions which are widely known as the
`DiVincenzo checklist':
\newline
i) One needs well-defined two-state quantum systems (qubits);
\newline
ii) One should be able to prepare the initial state of the qubits
with sufficient accuracy;
\newline
iii) A long phase time coherence is needed, sufficient to allow for
a large number ($\sim 10^4$) of coherent manipulations;
\newline
iv) Control over the qubit's Hamiltonian is required to perform the
necessary unitary transformations;
\newline
v) A quantum measurement is needed to read out the quantum
information.
\newline
At the present time any known physical system falls short of the lsit
stipulated above. Currently lots of different possibilities of building a
quantum computer are investigated including nuclear spins, quantum dots,
Josephson junctions, trapped ions, optical lattices, electrons on liquid
helium and some others with each of them being quite far from satisfying
this or that condition of the above set.

A route which seems to be  one of the most promising is to build a
system of Josephson junction (JJ) qubits: the qubits which are
based on Josephson junctions and utilise the charge and flux
degrees of freedom. The main advantages of the superconducting
devices involving Josephson junctions are a) a relative easiness
to manufacture - the lithographic methods  used to fabricate them
are well established; b) controllability - gate voltages
controlling charge Josephson qubits can be adjusted with a very
high degree of accuracy allowing, in particular, the high-fidelity
preparation of the initial state of the qubit; and finally c)
measurability - the techniques for almost noninvasive measurements
which can be used for this type of qubits are quite advanced. It
is point iii) in the DiVincenzo checklist - decoherence - which
causes the greatest worry for the  JJ qubit, as well as for any
other realisation of a solid state qubit. We stress again that
`spin up' and `spin down' states in solid state qubits are formed
by a substantial number of electrons (which is not the case for
nuclear spins which can be effectively decoupled from the external
world). All degrees of freedom interact with the environment thus
causing the loss of quantum coherence. Whether this problem of
decoherence can or cannot be satisfactorily solved is most likely
to determine the future of JJ qubits.

\begin{figure}
\centering
\includegraphics[height=4cm]{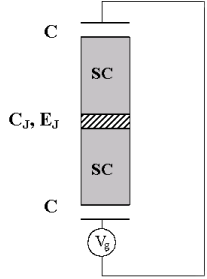}
\caption{Scheme for charge JJ qubit} \label{fig:qubitscheme}
\end{figure}

 The JJ qubit is made of
superconducting islands separated by a Josephson junction; the
scheme is depicted in Fig.\ref{fig:qubitscheme}.
In the scheme,  `SC' marks the superconducting islands and the
hatched area is the Josephson junction with capacity $C_J$ and
Josephson coupling $E_J$. The superconductors are separated from
the leads by capacitors $C$, which do not allow any tunneling, and
these capacitors are biased by some controllable gate voltage
$V_g$. Cooper pairs can tunnel through the Josephson junction thus
changing the total charge of each island.

\begin{figure}
\centering
\includegraphics[height=4cm]{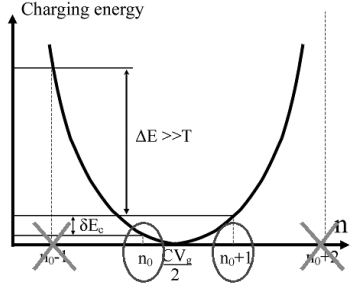}
\caption{Two charge states} \label{fig:coulombblockade}
\end{figure}

The charge JJ qubit is a two-level system which states are different by
the charge of a single Cooper pair. The equilibrium number of Cooper pair
in each island is controlled by the gate voltage.   When it is tuned to be
close to half integer $\sim n_0+1/2$ , the states with $n_0$ and $n_0+1$
Cooper pairs (circled in Fig.\ref{fig:coulombblockade}) can be arbitrarily
close to each other in energy while the distance to other energy levels
(crossed in Fig.\ref{fig:coulombblockade}) is of the order of the charging
energy that -- in the temperature units - could be tens or even a hundred
degrees of Kelvin which is much higher than typical experimental
temperatures. We assume that the Josephson coupling energy is much smaller
than the charging energy so that the states are really discriminated by
charge. (In the opposite limit, a so called flux JJ qubit can be -- very
successfully -- built). Thus we effectively separate all the charge states
with $n<n_0$ or $n>n_0+1$ and thus have a two level system, whose states
are made of a macroscopic number of electrons. In future we will be
referring to these as states `spin up' and `spin down'.

More formally,  the total energy of the system which consists of charging
and Josephson contributions can be written as follows:
\begin{align}
E=\frac{1}{2C_J+C}\left(2|e|n-\frac{CV_g}{2}\right)^2-E_J\cos\Theta,\label{energy}
\end{align}
where $\Theta$ is the superconducting order-parameter phase shift
between the islands. In properly chosen units, the pair of $n$ and
$\Theta$ are canonically conjugated coordinate and momentum since
they generate correct Hamiltonian equations of motion.
Differentiating with respect to momentum $\dd n/ \dd t=\partial
E/\partial\Theta=E_J\sin\Theta$ produces the correct equation for
the Josephson current (in the units chosen $E_J$ coincides with
the critical current), and differentiation with respect to
coordinate $\dd \Theta/\dd t=-\partial E/\partial
n=(CV_g-2n)/(2C_J+C)=V_J$ produces the correct value for the
voltage drop $V_J$ across the junction. The quantisation of
(\ref{energy}) then gives:
\begin{align}
\hat{H}=\frac{1}{2C_J+C}\left(\hat{n}-
\frac{CV_g}{2}\right)^2-E_J\cos\hat{\Theta},
\quad\quad\left[\hat{n},\hat{\Theta}\right]=\text{i}
.\label{hamiltonian-0}
\end{align}

After taking the projection of the Hamiltonian
(\ref{hamiltonian-0}) onto two states of interest, in this `spin
up' and `spin down' basis the charging part can be written as
$(\delta E_c/2)\hat{\sigma}_z$, where $\delta E_c$ is the charging
energy difference between the two states:
\begin{align}
\delta
E_c(V_g)=\frac{2}{2C_J+C}\left(n_0+\frac{1}{2}-\frac{CV_g}{2}\right)\,.\label{E_c}
\end{align}
The commutator between $\hat{n}$ and $\hat{\Theta}$ dictates the
commutation relations $[\hat{n},e^{\pm\text{i}\hat{\Theta}}]=\mp
e^{\pm\text{i}\hat{\Theta}}$ from which it immediately follows
that in the `spin up' and `spin down' basis $\cos\hat{\Theta}$ is
given by $\hat{\sigma}_x/2$. Summarizing the above we substitute
the Hamiltonian (\ref{hamiltonian-0}) with its projection:
\begin{align}\label{H-1}
\hat{H}=\frac{\delta
E_c(V_g)}{2}\,\hat{\sigma}_z-\frac{E_J}{2}\,\hat{\sigma}_x\,.
\end{align}

As the result we have a controllable two level system or a qubit.
The control is exercised by changing the gate voltage $V_g$ on
which the $\hat{\sigma}_z$ coefficient $\delta E_c$ depends
(\ref{E_c}).  If $\delta E_c$ is maintained large $\delta E_c\gg
E_J$, which is characteristic for charge JJ qubit, then the
Josephson part of the Hamiltonian is irrelevant and evolution
amounts to acquiring phase shift between `spin up' and `spin down'
states in the initial mixture. On the contrary, tuning charging
part to the degeneracy point $\delta E_c=0$ stimulates the spin
flip process. It can be shown that the ability to switch for
arbitrary time to the above two regimes is enough in order to
provide all necessary one qubit quantum operations or quantum
gates.

In building a quantum computer, one and two-qubit gates will be
required. The details of how to build them are beyond the scope of
these lectures and can be found in the comprehensive review
\cite{makhlin2} of Makhlin et. al. But what is crucial for any
quantum computation is to maintain entanglement between the
qubits. This will be undermined by inevitable decoherence.
Therefore, understanding its mechanisms and studying how it sets
in is crucial for future progress in this area.

\section{Decoherence - Fluctuating Background Charges model}
\label{sec:fbcmodel}
\subsection{The model}
The main problem with charge JJ qubits (as with any other solid
state qubits) is that they lose quantum coherence too quickly due
to unwanted, but unavoidable, coupling to the environment. Recent
experiments [11--15]
show that
hundreds of elementary quantum operations can be achieved before
coherence is destroyed. This is at least two orders of magnitude
short \cite{loss3} of a rough estimate of what is required for
non-trivial quantum calculations. Thus the problem of decoherence
is currently the major obstacle for the progress of qubits of this
type, and requires a close theoretical examination.

Spin-boson models will not offer us any insight into how real
physical parameters of the problem affect decoherence. To obtain
this we   need a physical model describing the real processes
which decohere the qubit. It is widely believed
\cite{nakamura2,pashkin} that in charge JJ qubits the main
contribution to decoherence comes from coupling of the qubit to
some charge fluctuators present in the environment. The
fluctuating background charges (FBC) model, suggested in this
context in \cite{paladino1} and then also used in
\cite{galperin1,galperin2,gyl2,faoro}, is an attempt to build a
microscopic model for such a dominant channel of decoherence.

\begin{figure}
\centering
\includegraphics[height=4cm]{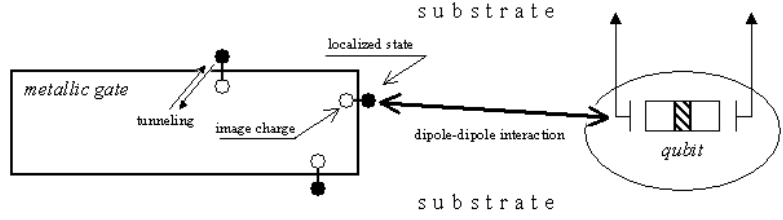}
\caption{Interaction between the qubit and fluctuating background
charges} \label{fig:fbc}
\end{figure}

The schematic picture of how the qubit interacts with the FBC is
shown in Fig.\ref{fig:fbc}.
Impurities which contribute to decoherence  sit on the substrate
which the nanocircuit was grown on, close enough to a metallic
lead to allow tunneling to and from the lead. Solid circles in
Fig.\ref{fig:fbc} are charged impurities and transparent circles
are their image charges. These dipoles interact electrically with
the qubit, which in turn behaves like a dipole, since one of the
superconducting islands has some number of excessive Cooper pairs
and the other one lacks the same number. The interaction depends
on the number of such extra pairs and thus is different for the
two charge states of the qubit, enforcing dependence of the
coupling term on $\hat{\sigma}_z$. Although a static dipole-dipole
interaction could only shift the qubit states without any loss of
coherence, the tunneling between the impurities and the metallic
lead makes charges on the impurities to fluctuate, effectively
creating a random time-dependent field on the qubit. This random
field  causes decoherence. This is similar to the generic case
considered in Section \ref{sec:dec}: the time-dependent electric
field coupled to $\sigma_z$ plays the role of the random `magnetic
field' $B_z$ in Eq.~(\ref{rho-av}). In the following section, we
will show that within the Fermi golden rule approximation (valid
for a weak coupling between the qubit and the FBC), both
decoherence and relaxation can be calculated in a way similar to
that schematically developed in Section \ref{sec:dec}, with an
additional advantage of calculating the noise power-spectrum
within a fully microscopical model. In this section, we restrict
considerations to the  longitudinal model of pure decoherence. Its
advantage is that one can go well beyond the Fermi golden rule,
and calculate the decoherence rate (but not the relaxation!)
non-perturbatively and within a fully quantum description valid
for any temperature, including $T\to0$ (in which limit there is no
decoherence in the absence of relaxation).

The Hamiltonian corresponding to the described model   can be
written as
\begin{align}
\hat{H}&=\frac{\delta
E_c}{2}\,\hat{\sigma}_z-\frac{E_J}{2}\,\hat{\sigma}_x+
\hat{\sigma}_z\hat{V}+\hat{H}_{\text
B};\quad\quad\hat{V}=\frac{1}{2}
\sum\limits_iv_i^{\phantom+}\!\hat{d}^\dagger _i
\hat{d}^{\phantom+}_i;\nonumber\\
\hat{H}_{\text B}&=\sum\limits_i\varepsilon^0_i \hat{d}^\dagger
_i\hat{d}^{\phantom+}_i+
\sum\limits_{i,\textbf{k}}\left[t_{\textbf{k}i}^{\phantom+}
\!\hat{c}^\dagger _{\textbf{k}}\hat{d}^{\phantom+}_i+ \text{
h.c}\right]+\sum\limits_{\textbf{k}}\varepsilon_{\textbf{k}}
^{\phantom+}\!\hat{c}^\dagger _{\textbf{k}}\hat{c}
^{\phantom+}_{\textbf{k}} .\label{themodel}
\end{align}
Here $\hat{d}^{\phantom+}_i$, $\hat{d}^\dagger _i$ are the
operators of annihilation and creation of an electron on the i-th
impurity; $\hat{c}^{\phantom+}_{\textbf{k}}$, $\hat{c}^\dagger
_{\textbf{k}}$ are the operators of the conduction electrons in
the metal; $t_{\textbf{k}i}$ are the hybridisation amplitudes;
$\varepsilon^0_i$ is the energy of the localised state on the i-th
impurity; $\varepsilon_{\textbf{k}}$ are the energies of the
conducting electrons; $v_i$ is the coupling strength between the
qubit and the $i^{{\text{th}}}$ impurity.

The charge JJ qubit corresponds to $\delta E_c\gg E_J$, which is always
the case for the small enough capacities (i.e.\ small superconducting
islands and the JJ junction in Fig.~\ref{fig:qubitscheme}.  A non-trivial
circuit is supposed to consist of many qubits, and most of the time each
particular qubit is in an idle regime for which the condition above is
fulfilled. During relatively short times necessary for operations
involving spin-flips the longitudinal model, $E_J=0$, is not appropriate.
However, if there is any hope for a working qubit, the onset of
decoherence should only happen in the idle regime.

Apart from temperature, there are three other parameters of
dimensions of energy  for each  fluctuator: the coupling strength
$v_i$, tunneling rate $\gamma_i$
($\gamma_i=2\pi\sum_{\textbf{k}}|t_{\textbf{k}i}|^2\delta(\omega-
\varepsilon_{\textbf{k}})$), and fluctuator energy $\varepsilon_i$
(note that the bare energy $\varepsilon_i^0$ is renormalised by
hybridisation). All three parameters are broadly distributed, and
temperature can be considered `low' for some of them and `high'
for  others. In the Section \ref{sec:main} we will show that these
two regimes are defined as follows:
\begin{align}
\left\{
\begin{array}{lcl}
T\ll\text{min}\left|\varepsilon_i\pm\frac{1}{2}\sqrt{v^2_i-\gamma_i^2}\right|&
\text{or}\quad T\ll\gamma_i
& \quad\quad\text{low-T regime}\\
\phantom.\\
T\gg\text{max}\left|\varepsilon_i\pm\frac{1}{2}\sqrt{v^2_i-\gamma_i^2}\right|&
\text{and}\quad T\gg\gamma_i
& \quad\quad\text{high-T regime}\\
\end{array}\right.\label{temp-class}
\end{align}

In the high-temperature regime, the decoherence rate can be
calculated classically \cite{paladino1,galperin1} which gives for
one fluctuators (omitting the index $i$)
\begin{align}
\Gamma_2=\frac{\gamma}{2}\left[1-\Re\text{e}
\sqrt{1-\frac{v^2}{\gamma^2}}\right]=
\left\{
\begin{array}{ll}
\gamma/2\,, & v>\gamma\\
\phantom.\\
 v^2/4\gamma\,, & v\ll\gamma
\end{array}\right.
\,.\label{classical-gamma2}
\end{align}
The high temperature treatment of the model would be justified if
the high energy impurities,  $|\varepsilon_i|\gtrsim T$,  were
frozen. However, the same hybridisation $\gamma$ which makes
impurity charges to fluctuate, broadens their energy positions.
Such a broadening creates a Lorentzian tail which is, of course,
power-law suppressed but nevertheless allows for a contribution
from the energetically remote impurities. Moreover, even
impurities with $|\varepsilon_i|\ll  T$ may not satisfy the
high-$T$ inequality  (\ref{temp-class}) and contribute in a
non-classical way. We will present below some arguments that the
impurities most relevant for decoherence  are likely to to be in
the low-temperature regime.

\subsection{Why low temperature?}

Let us first illustrate why the hybridisation leads to the
Lorentzian tail. To simplify notation,  we will be considering a
single impurity. The quadratic Hamiltonian of the bath
$\hat{H}_{\text B}$ can be diagonalised by means of some linear
transformation \cite{mahan}, after which it becomes
$\hat{H}_{\text B}=\sum_n\epsilon_n\hat{\alpha}^\dagger
_n\hat{\alpha}^{\phantom+}_n$. The interaction part $\hat{V}$ gets
transformed into a sum over all exact states:
$\hat{V}=(v/2)\sum_{n,m}u^*_nu^{\phantom*}_m\hat{\alpha}^\dagger
_n\hat{\alpha}^{\phantom{+}}_m$, where coefficients $u_n$ are
related to the density of states (DoS) $\nu_{\varepsilon}(\omega)$
on the impurity broadened by hybridisation with the conduction
band: $|u_n|^2=\nu_{\varepsilon}(\epsilon_n)\delta$; here $\delta$
is the level spacing in the conduction band, and the DoS   is
given by the Lorentzian
\begin{align}
\nu_{\varepsilon}(\omega)=\frac{1}{\pi}
\frac{\gamma/2}{(\omega-\varepsilon)^2
+\gamma^2/4}\,.\label{lorentzian}
\end{align}
\begin{figure}
\centering
\includegraphics[height=4cm]{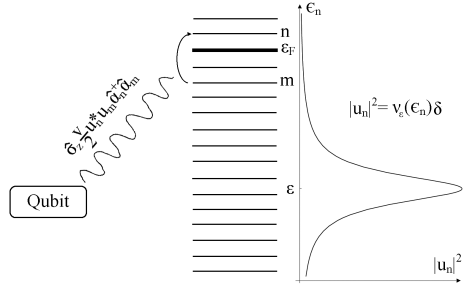}
\caption{Qubit is coupled to the transitions between all exact
states of the bath Hamiltonian $\hat{H}_{\text B}$, with weight
being proportional to $u^*_nu_m$. Coefficients $u^*_n, u_m$ are
governed by the Lorentzian
$|u_n|^2=\nu_{\varepsilon}(\epsilon_n)\delta$.} \label{fig:dos}
\end{figure}
In this notation, low temperature manifests itself in smallness of
the interaction term $\hat{V}$, to which only $n$ and $m$ in the
vicinity of the Fermi level will effectively contribute (see
Fig.\ref{fig:dos}). Since the smallness of $u$ coefficients is
controlled by the Lorentzian (\ref{lorentzian})(rather than
distribution function), the effect of such a high energy impurity
can only be power law suppressed.

The second observation concerns very stringent geometrical
restrictions on where an `effective' impurity can be and a broad
distribution of energies of impurities. In order to allow tunneling
to and from the impurity, it should not be too far away from the
metallic gate, which implies that the total volume available for
`effective' impurities cannot exceed $1\mu m\times(10\lambda_F)^2$,
where $1\mu m$ is a characteristic size of a metallic gate and
$\lambda_F$ is the Fermi wavelength in the metal. If we are hoping
to find at least one low energy $|\varepsilon_i|< T$ impurity in
this volume, the total number of impurities there should be of order
of $D/T$, where $D$ is the width of the distribution function of
energies. For temperature which varies in experiments between $30mK$
and $50mK$, and for $D$ having a typical chemical value of $1eV$ the
above ratio is of order of $10^5$. Having that many impurities in a
volume which in total accommodates $\sim 10^6$ atoms is completely
unrealistic.

So, the contribution of high energy impurities is suppressed but not
strongly enough in order to exclude them from consideration, and low
energy impurities effectively do not exist. If for a particular
sample there happened to be one, this sample will exhibit orders of
magnitude stronger decoherence, and should better be discarded.

The condition $|\varepsilon_i|\gg T$ enforces low temperature regime
which requires a full quantum mechanical treatment. Such a
treatment, valid at any temperature, will be offered in the next
Section.

\section{Exact solution for FBC model at arbitrary temperature}
\label{sec:main}
\subsection{Calculations}

To calculate the decoherence rate we will need to know how the
off-diagonal elements of the reduced density matrix of the qubit
decay with time. The dynamics of the full density matrix of the
whole system qubit + bath is given by the Heisenberg equation of
motion:
\begin{align}
\frac{\partial\hat{\rho}(t)}{\partial
t}=-\text{i}\left[\hat{H},\hat{\rho}(t)\right]\,,\label{h-comm}
\end{align}
where $\hat{H}$ is the Hamiltonian given by the Eq.
(\ref{themodel}) with Josephson energy being taken equal to zero.
Since in the longitudinal ($E_J=0$) model the interaction term
commutes with the qubit Hamiltonian, one can trace out the bath
degrees of freedom thus yielding the following result for the
reduced density matrix of the qubit:
\begin{align}
\hat{\rho}^{\,(q)}(t)=\,\left(
\begin{array}{cc}
\rho^{(q)}_{11}(0) & \rho^{(q)}_{12}(0)\,e^{-\text{i}\delta E_ct}D(t)\\
\rho^{(q)}_{21}(0)\,e^{\,\text{i}\delta E_ct}D^*(t) & \rho^{(q)}_{22}(0)\\
\end{array}\right)\,,\label{rdm}
\end{align}
where the standard separable initial condition for the full
density matrix,
$\hat{\rho}(0)=\hat{\rho}^{\,(q)}(0)\otimes\hat{\rho}_B;\,\hat{\rho}_B
=\ee^{-\beta\hat{H}_{\text B}}/\Tr\ee^{-\beta\hat{H}_{\text B}}$,
was assumed. The diagonal elements of the reduced density matrix
do not evolve since there are no spin flip processes present, and
the dynamics of the off-diagonal elements is governed by the
decoherence function $D(t)$:
\begin{align}
D(t)=\left\langle e^{\,\text{i}(\hat{H}_{\text B}+\hat{V})t}
\ee^{-\text{i}(\hat{H}_{\text
B}-\hat{V})t}\right\rangle\,,\label{main-average}
\end{align}
where averaging should be performed with the bath Hamiltonian
$\hat{H}_{\text B}$. At long times $t$  the decoherence function
must decay  exponentially, $D(t)\sim \ee^{-\Gamma_2t}$, so that
the decoherence rate $\Gamma_2$ is defined as
\begin{align}
\Gamma_2=-\Re\text{e}\lim\limits_{t\rightarrow\infty}t^{-1}\ln
D(t)\,,\label{def-dec}
\end{align}
in the agreement with Eq.~(\ref{G-formal}).

\begin{figure}[t]
\vspace*{2cm}

\centering
\includegraphics[clip=true,height=4cm]{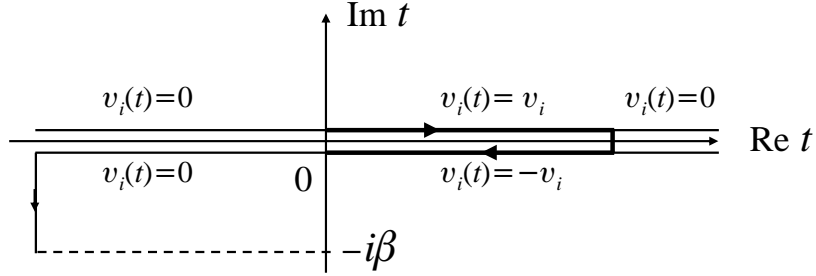}
\caption{Dependence of $v_i(t)$ on time along the Keldysh contour}
\label{fig:keldysh}
\end{figure}

The average in Eq.~(\ref{main-average}) can be represented as the
following functional integral with the Grassmann fields $\xi$ and
$\eta$ defined on the Keldysh contour:
\begin{align}
D(t)=Z^{-1}\int &
\mathcal{D}\xi^*\mathcal{D}\xi\mathcal{D}\eta^*\mathcal{D}\eta\,
\ee^{\phantom{\int\limits_c}\!\!\!\text{i}S_0\left[\xi\right]}
\ee^{\phantom{\int\limits_c}\!\!\!\text{i}S_0\left[\eta\right]}
\ee^{ \frac{\text{i}}{2}\!\int\limits_{c_K}\sum\limits_{i}
v_i(t')\xi^*_i(t')\xi^{\phantom*}_i(t')\dd t'}\times\nonumber\\
& \times \ee^{-\text{i}\int\limits_{c_K}\sum\limits_{{\bf
k},i}\left[t^{\phantom*}_{{\bf k}i}\eta^*_{\bf
k}(t')\xi^{\phantom*}_i(t')+ t^*_{{\bf
k}i}\xi^*_i(t')\eta^{\phantom*}_{\bf k}(t')\right]\dd
t'}\,.\label{fint}
\end{align}
Here $Z$ is the same functional integral but with $v_i(t')\equiv
0$, the fields $\xi^{\phantom*}_i,\,\xi^*_i$ correspond to the
localised state on the i-th impurity, $\eta^{\phantom*}_{\textbf
k},\,\eta^*_{\textbf k}$ to the conduction electrons, the impurity
$S_0[\xi]$ and free electrons $S_0[\eta]$ actions are given by:
\begin{align}
& S_0\left[\xi\right]=\int\limits_{c_K}\sum\limits_{i}\xi^*_i(t')
(\text{i}\partial_{t'}-\varepsilon^0_{i})\xi^{\phantom*}_i(t')\dd t'\nonumber\\
& S_0\left[\eta\right]=\int\limits_{c_K}\sum\limits_{\bf
k}\eta^*_{\bf k}(t')(\text{i}\partial_{t'}-\varepsilon_{\bf
k})\eta^{\phantom*}_{\bf k}(t')\dd t'\,,\label{0-actions}
\end{align}
and the Keldysh time dependent coupling $v_i(t')$ is zero everywhere
on the contour apart from the interval $(0,t)$ where on the upper
branch it takes the value of $v_i$, and on the lower branch it is
equal to $-v_i$ (see Fig.\ref{fig:keldysh}).
Since the action $S_0[\eta]$ is quadratic and the hybridisation
term is linear in $\eta$ fields, it is straightforward to
integrate out the conduction electrons:
\begin{align}
D(t)=Z^{-1}_0\int\mathcal{D}\xi^*\mathcal{D}\xi\,
e^{\phantom{\int\limits_c}\text{i}S_0\left[\xi\right]}&
e^{-\text{i}\iint\limits_{c_K}\sum\limits_{ij}\xi^*_i(t_1)\Sigma_{ij}
(t_1,t_2)\xi^{\phantom*}_j(t_2)\dd t_1\dd t_2}\nonumber\\
&\quad\quad\quad\quad\quad e^{\phantom{.}\frac{\text{i}}{2}
\int\limits_{c_K}\sum\limits_{i}v_i(t_1)\xi^*_i(t_1)\xi^{\phantom*}_i(t_1)\dd
t_1}\label{after-int1}
\end{align}
Again, here $Z$ is the same functional integral but with
$v_i(t')\equiv 0$. All the information about the conduction
electrons is now encoded in the self-energy matrix which is
defined on the contour:
\begin{align}
\Sigma_{ij}(t_1,t_2)=\sum\limits_{\bf k}t^*_{{\bf
k}i}t^{\phantom*}_{{\bf k}j}g_{\bf k}(t_1,t_2)\,,\quad\quad
t_1,t_2\in c_K\label{selfenergy-matrix}
\end{align}
with $g_{\bf k}(t_1,t_2)$ being the Green function of free
electrons. The full action in (\ref{after-int1}) is quadratic and
therefore the corresponding functional integral can be written in a
standard symbolic trace-log notation:
\begin{align}
D(t)=e^{\,\text{Tr}\ln\left[\hat{1}+\hat{G}\hat{U}\right]}\,,
\label{trace-log}
\end{align}
where both $\hat{G}$ and $\hat{U}$ are matrices in the space of
impurity indices, and both depend on two times along the Keldysh
contour. The symbol $\Tr$ here stands for trace with respect to
the impurity indices and for integration along the Keldysh contour
over all the times involved. The Green function $\hat{G}$ obeys
the following integro-differential equation:
\begin{align}
\int\limits_{c_K}\dd
t'\sum\limits_{m}\left[\delta_{im}\delta(t_1,t')
\left(\text{i}\frac{\partial}{\partial
t'}-\varepsilon^0_m\right)-\Sigma_{im}(t_1,t')\right]G_{mj}(t',t_2)
=\delta_{ij}\delta(t_1,t_2)\,,\label{green-int-diff}
\end{align}
with $\delta$-function $\delta(t_1,t_2)$ being defined on the
contour. The impurity matrix $\hat{U}$ is made out of the
time-dependent coupling $v_i(t')$:
\begin{align}
U_{ij}(t_1,t_2)=\delta_{ij}\delta(t_1,t_2)\frac{v_i(t_2)}{2}\,.
\end{align}
Being written explicitly, the symbolic expression in
(\ref{trace-log}) becomes
\begin{align}
D(t)=e^{-\sum\limits^{\infty}_{n=1}\frac{(-1)^n}{n}\,\,
\text{tr}\idotsint\limits_{\substack{c_K
\\\mbox{\tiny 2n integrals}}}\dd t_1\dd t'_1\ldots \dd t_n\dd t'_n
\hat{G}(t_1,t_1')\hat{U}(t_1',t_2)\ldots
\hat{G}(t_n,t_n')\hat{U}(t_n',t_1)}\label{expansion1}
\end{align}
where tr stands for a trace over impurity indices only.

Since the matrix $\hat{U}(t_1,t_2)$ as function of its both times
is non-vanishing  only between $0$ and $t$, all the contour
integrals in (\ref{expansion1}) can be converted into ordinary
integrals from $0$ to $t$  by introducing the Keldysh structure
for both matrices $\hat{G}$ and $\hat{U}$. Both arguments of
$\hat{G}(t_1,t_2)$ can either be on the upper or on the lower
branch of the Keldysh contour leaving four choices which generate
the following Keldysh structure:
\begin{align}
&G_{ij}(t_1,t_2)_{ t_{1,2}\in c_K}\mapsto
 \check{\mathcal{G}}_{ij}(t_1-t_2)_{ t_{1,2}\in
(-\infty,+\infty)}\!=\!
\begin{pmatrix}
\mathcal{G}_{ij}(t_1-t_2) & \mathcal{G}^<_{ij}(t_1-t_2)\\[4pt]
\mathcal{G}_{ij}^>(t_1-t_2) &
\widetilde{\mathcal{G}}_{ij}(t_1-t_2)\\
\end{pmatrix}\label{gcheck}
\end{align}
where ${\mathcal{G}}_{ij}(t_1-t_2)$ and
$\widetilde{\mathcal{G}}_{ij}(t_1-t_2)$ are the time-ordered and
anti-time-ordered  Green functions. Using the Keldysh structure
for the $\hat{U}$-matrix,
\begin{align*}
U_{ij}(t_1,t_2)_{ t_{1,2}\in c_K}\mapsto\tfrac12
\delta_{ij}\,\hat{\sigma}_z \,\delta(t_1-t_2)\,{v_i(t_2)}_{
t_{1,2}\in (-\infty,+\infty)}\,,
\end{align*}
and the general rule,
\begin{align*}
\int\limits_{c_K}A(t_1,t')B(t',t_2)&\dd t'\bigg|_{ t_{1,2}\in
c_K}\mapsto  \int\limits_{-\infty}^{+\infty}\check{A}
(t_1,t')\hat{\sigma}_z\check{B}(t',t_2)\dd t' \bigg|_{  t_{1,2}\in
(-\infty,+\infty)}\,,
\end{align*}
we rewrite the expression (\ref{expansion1}) for the decoherence
function as follows:
\begin{align}
D(t)=e^{-\sum\limits^{\infty}_{n=1}\frac{(-1)^n}{n2^n}\,\,\text{Tr}
\int\limits_0^t\ldots\int\limits_0^t \dd t_1\dd t_2\ldots \dd t_n
\check{\mathcal{G}}(t_1-t_2)(\hat{v}\otimes\hat{1})
\check{\mathcal{G}}(t_2-t_3)(\hat{v}\otimes\hat{1})\ldots
\check{\mathcal{G}}(t_n-t_1)(\hat{v}\otimes\hat{1})}\label{expansion3}
\end{align}
Now the symbol Tr stands for the trace over both  impurity and
Keldysh matrix indices, while $(\hat{v}\otimes\hat{1})$ denotes
the matrix with elements $v_i$ on the main diagonal  in the
impurity space and which is the unity matrix in the Keldysh space.

It is convenient to do a standard rotation \cite{rammer} in
Keldysh space:
\begin{align}
\hat{\mathcal{G}}_{ij}=\hat{L}\hat{\sigma}_z\check{\mathcal{G}}_{ij}
\hat{L}^\dagger ;\quad\quad
\hat{L}=\frac{1}{\sqrt{2}}(\,\hat{1}-\text{i}\hat{\sigma}_y)\,,
\end{align}
which converts the four-element  matrix in (\ref{gcheck}) into the
three-element one of retarded, advanced and Keldysh Green
functions:
\begin{align}
\hat{\mathcal{G}}_{ij}=
\begin{pmatrix}
G^R_{ij} &   & G^K_{ij}\\[6pt]
0 &   & G^{A}_{ij}\\
\end{pmatrix} \end{align}
In the rotated basis the expression (\ref{expansion3}) for the
decoherence function becomes:
\begin{align}
D(t)=e^{-\sum\limits^{\infty}_{n=1}\frac{(-1)^n}{n2^n}\,\,\text{Tr}
\int\limits_0^t\ldots\int\limits_0^t \dd t_1\dd t_2\ldots \dd t_n
(\hat{v}\otimes\hat{\sigma}_x)\hat{\mathcal{G}}(t_1-t_2)
(\hat{v}\otimes\hat{\sigma}_x)\hat{\mathcal{G}}(t_2-t_3)\ldots
(\hat{v}\otimes\hat{\sigma}_x)\hat{\mathcal{G}}(t_n-t_1)}\label{expansion4}
\end{align}
Since due to the time translation invariance each $\hat G$ depends
only on the difference of its time arguments, the
$n^{{\text{th}}}$ order integrand depends on $n-1$ differences in
times, while the integration over the last time variable produces
the overall factor of $t$. The region of integration becomes, at
arbitrary time, quite complicated. However, when time t is much
bigger than the characteristic time on which
$\hat{\mathcal{G}}(\tau)$ function decays (which is true if
$t\gg\gamma^{-1}_i,T^{-1}$), all the  integrals over the time
differences can be extended to the entire axis.  Then the integral
has a convolution structure in time and, performing a Fourier
transform and a straightforward summation that restores the
logarithm, it finally reduces to
\begin{align}
D(t)=
\ee^{t\int\limits^{+\infty}_{-\infty}\frac{\dd\omega}{2\pi}\Tr
\ln\left[ \hat{1}+\frac{1}{2}(\hat{v}\otimes\hat{\sigma}_x)
\hat{\mathcal{G}}(\omega)\right]}\,. \label{resummation}
\end{align}
Using the definition of decoherence rate (\ref{def-dec}) and
calculating trace in the Keldysh space explicitly we have:
\begin{align}
\Gamma_2=-\Re\text{e}\int\limits^{+\infty}_{-\infty}
\frac{\dd\omega}{2\pi}\,\operatorname{tr}\ln\left[
\hat{1}+\frac{\hat{v}}{2}\,
\hat{G}^K(\omega)-\frac{\hat{v}}{2}\,\hat{G}^R(\omega)
\frac{\hat{v}}{2}\,\hat{G}^A(\omega)\right]\,,
\label{decoherence-manyfluct}
\end{align}
where $\operatorname{tr}$ refers only to the impurity matrix
indices.

It should be stressed that the analytical structure of the
expression in the r.h.s.\ of Eq.~(\ref{decoherence-manyfluct})
ensures that there is no decoherence at zero temperature. Indeed,
using the standard relation  \cite{rammer} between the components
of Keldysh Green functions at equilibrium,
$\hat{G}^K(\omega)=\tanh(\beta\omega/2)
[\hat{G}^R(\omega)-\hat{G}^A(\omega)]$, which at $T=0$ reduces to
$\hat{G}^K(\omega)=\text{sgn}(\omega)[\hat{G}^R(\omega)-\hat{G}^A(\omega)]$,
we have:
\begin{align}
\Gamma_2(T=0)=-\Re\text{e}\int\limits^{+\infty}_{-\infty}
\frac{\dd\omega}{2\pi}\,\text{tr}\ln\left[
\left(\hat{1}+\frac{\hat{v}}{2}\,\hat{G}^R(\omega)\right)
\left(\hat{1}-\frac{\hat{v}}{2}\,\hat{G}^A(\omega)\right)\right]
\,.\label{zero-temp-manyfluct}
\end{align}
Separating the retarded and advanced parts in the expression
above, and then expanding the logarithm in both of them, we see
that starting from the second order all the integrals are
convergent at $\omega\to\infty$ and therefore equal to zero since
the retarded (advanced) Green function is analytic in the upper
(lower) half plane. Combining the first order contributions from
both parts we see that it is purely imaginary and therefore also
vanishes.

Assuming that the impurity states are well localised, the
tunneling amplitudes $t_{{\bf k}i}$ can be written as $t_{{\bf
k}i}=t_i\text{Vol}^{-1/2}e^{\,\text{i}{\bf k}{\bf r}_i}$, where
${\bf r}_i$ is the position of the i-th impurity. This in turn
means that the self energy matrix (\ref{selfenergy-matrix}) has
the following representation:
\begin{align}
\hat{\Sigma}_{ij}(t_1-t_2)=t^*_it^{\phantom*}_j\hat{g}(t_1-t_2;{\bf
r}_j-{\bf r}_i)\,.
\end{align}
Since the characteristic distance $\it{l}$ between the impurities
is much bigger than the Fermi wavelength, $\it{l}\gg\lambda_F$,
all the off-diagonal elements of the self energy matrix will be
averaged out while diagonal can be represented as
\begin{align}
\Sigma^{A/R}_{ii}(\omega)= \left[\sum\limits_{\bf k}\frac{|t_{{\bf
k}i}|^2}{\omega-\varepsilon_{\bf k}}\pm\text{i}\pi\sum\limits_{\bf
k}|t_{{\bf k}i}|^2\delta(\omega-\varepsilon_{\bf k})\right]=
\left[\frac{\alpha_i(\omega)}{2}\pm\text{i}
\frac{\gamma_i(\omega)}{2}\right]\label{real-im}\,.
\end{align}
This allows one to solve (the Keldysh rotated analogue of)
Eq.~(\ref{green-int-diff}):
\begin{align}
G^{A/R}_{ij}(\omega)=
\frac{\delta_{ij}}{\omega-\varepsilon^0_i-\Sigma^{A/R}_{ii}(\omega)}
=\frac{\delta_{ij}}
{\omega-\varepsilon_i\mp\text{i}\gamma_i/2}\,,\label{ret-adv}
\end{align}
where both the real and imaginary parts of the self energy matrix
were assumed $\omega$-independent, and the real part was absorbed
by renormalisation of the energy levels
$\varepsilon_i=\varepsilon^0_i+\alpha_i/2$. In such a  diagonal
approximation for the self-energy, the contributions from all the
impurities are independent and the total decoherence rate is given
by their sum. Substituting Eq.~(\ref{ret-adv}) for the Green
functions into Eq.~(\ref{decoherence-manyfluct}), subtracting (to
improve the integral convergency) the identically zero expression
(\ref{zero-temp-manyfluct}) for ${\Gamma}_0$ and taking the real
part of the resulting expression, we obtain the following
contribution of a single impurity at energy ${\varepsilon} _j
\equiv {\varepsilon}$ to the decoherence rate:
\begin{align}
\Gamma_2(T)&=
-\int\limits^{+\infty}_{-\infty}\frac{\dd\omega}{4\pi}\ln\left[1-
\frac{v^2\gamma^2\cosh^{-2}(\omega/2T)}
{v^2\gamma^2+4\big[(\omega-\varepsilon)^2+
\gamma^2/4-v^2/4\big]^2}\right]\,.\label{modulus}
\end{align}

\subsection{Discussion of the results}

It is convenient to introduce an auxiliary `spectral function'
$\Lambda(\omega)$,
\begin{align}\label{L}
    \Lambda(\omega)=\frac{1}{1+4(\omega-\varepsilon_+)^2
    (\omega-\varepsilon_-)^2/v^2\gamma^2}\,,\quad\quad\quad
\varepsilon_{\pm}=\varepsilon\pm\frac{1}{2}\sqrt{v^2-\gamma^2}\,,
\end{align}
thus rewriting Eq.~(\ref{modulus}) as follows:
\begin{align}
&\Gamma_2(T)=
-\int\limits_{-\infty}^{+\infty}\frac{\dd\omega}{4\pi}\ln\Bigg[
1-\Lambda(\omega)\cosh^{-2}\frac{\omega}{2T}\Bigg]
\label{gamma2}\,.
\end{align}

Behaviour of $\Lambda(\omega)$ is qualitatively different
depending on whether $g\equiv v/\gamma<1$ or $g>1$. In the former
case $\Lambda(\omega)$ is peaked at $\omega=\varepsilon$, while in
the latter the peak splits into two Lorentzians which are centered
at $\omega=\varepsilon_{\pm}$ and travel in different directions
away from the $\omega=\varepsilon$ point as $v$ increases. This
signals a qualitative change between the weak coupling ($g\ll 1$)
and
 the strong coupling ($g\gg 1$) regimes.

The decoherence rate is determined by the overlap of the
temperature function, $\cosh^{-2}(\omega/2T)$, and
$\Lambda(\omega)$ in (\ref{gamma2}). The former is  centered at
$\omega=0$ and exponentially decays for  $|\omega|>T$, while the
latter has the power-law decay away from the above described
peaks. If the temperature function is wide and the Lorentzian
peaks are sitting inside $T$, the overlap is solely controlled by
$\Lambda(\omega)$ as if the temperature function was equal to $1$
everywhere. In the opposite case when the temperature function is
narrow either with respect to the distance to the
$\Lambda$-function peaks or to their width, the overlap is
controlled by $\cosh^{-2}(\omega/2T)$ and $\Lambda(\omega)$ can be
replaced by $\Lambda(\omega=0)$. We will be calling these two
cases high-temperature
($T\gg\text{max}|\varepsilon_{\pm}|,\gamma$) and low-temperature
($T\ll\text{min}|\varepsilon_{\pm}|\,\,\text{or}\,\,T\ll\gamma$)
regimes correspondingly.

\begin{figure}
\centering
\includegraphics[height=4cm]{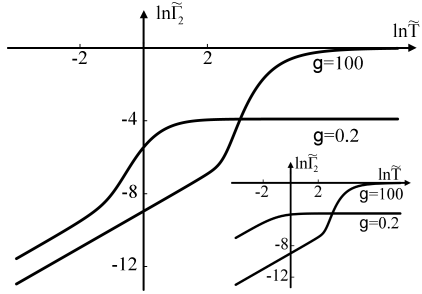}
\caption{Dependence of the decoherence rate on temperature,
Eq.(\ref{gamma2}), for strong and weak coupling. The main picture
shows a fluctuator with $\nu_{\varepsilon}(\omega)$ centered at
$\tilde{\varepsilon}=3$, the insert shows $\tilde{\varepsilon}=0$.
$\tilde{T},\tilde{\Gamma}_2,\tilde{\varepsilon}$ are measured in
units of $\gamma/2$.} \label{fig:tdep}
\end{figure}

In the high-temperature regime,
$T\gg\text{max}|\varepsilon_{\pm}|\,,\gamma$, the decoherence rate
$\Gamma_2$ saturates \cite{paladino1,galperin1} at
$\varepsilon$-independent value:
\begin{align}
\Gamma_2^{{\text{(high)}}}=
-\int\limits_{-\infty}^{+\infty}\frac{\dd\omega}{4\pi}\ln\left[
1-\Lambda(\omega)\right]=\frac{\gamma}{2}
\left[1-\Re\text{e}\sqrt{1-g^2}\right]\,.
\end{align}
In the low-T regime, when either
$T\ll\text{min}|\varepsilon_{\pm}|\,\,\text{or} \,\,T\ll\gamma$,
the decoherence rate becomes linear in temperature:
\begin{align}
\Gamma_2^{{\text{(low)}}}(T)&=-\int\limits_{-\infty}^{+\infty}
\frac{\dd\omega}{4\pi}\ln\Bigg[
1-\Lambda(\omega=0)\cosh^{-2}\frac{\omega}{2T}\Bigg]\nonumber\\[6pt]
&=\frac{T}{\pi}\arctan^2\left(\frac{2g}
{4\varepsilon^2/\gamma^2-g^2+1}\right) \,.\label{tlow}
\end{align}
When both the coupling strength $v$ and energy $\varepsilon$ are
smaller than the tunneling rate $\gamma$, the regime change occurs
in the vicinity of $T\sim\gamma$ point (the upper curve on the
insert in Fig.\ref{fig:tdep}). If $\gamma$ is smaller than either
of the other two parameters: $\gamma\ll
A=\text{max}\{\varepsilon,v/2\}$, then the high- and
low-temperature asymptotes parametrically mismatch. In this case
there is a crossover between the two regimes which is
exponentially fast and occurs in the logarithmically narrow
interval of temperatures:
\begin{align}
\Gamma_2^{{\text{(x-over)}}}(T)\sim\Gamma_2^{{\text{(high)}}}
\ee^{-A/T}\,,\quad\quad
\frac{A}{\ln A/\gamma}\lesssim T\lesssim A\,.
\end{align}
The temperature dependence of the decoherence rate is shown in
Fig.\ref{fig:tdep} for the case of different couplings and
fluctuator energies.
We should stress that although a linear in $T$ behaviour similar
to that in the Eq.(\ref{tlow}) would also follow from the
spin-boson models with the ohmic spectral function, as briefly
described in Section \ref{sec:dec}, only a full quantum mechanical
treatment can result in a non-trivial $T$ dependence depicted in
Fig.\ref{fig:tdep}.

Let us also stress that, although that at high temperatures
decoherence is always stronger for a strongly coupled fluctuator,
 this does not have to be the case as temperature is lowered,
since in the crossover region the $g\gg 1$ fluctuators undergo
stronger suppression than their $g\ll 1$ counterparts. Such a
non-monotonic dependence of the decoherence rate $\Gamma_2$ on the
coupling strength $v$ outside the classical region of high $T$
deserves special attention.

When we start with a small value of $v$, $v<\gamma$, the peak of
$\Lambda(\omega)$ grows as $v$ increases which leads to higher
decoherence rate $\Gamma_2$. This proceeds on until the $v=\gamma$
point is reached when the peak stops growing but splits into two
instead. One of those peaks then starts moving towards the origin
which increases the overlap between the temperature function and
$\Lambda(\omega)$, so $\Gamma_2$ keeps on growing. The maximum
possible overlap is achieved when the incoming Lorentzian is
centered at $\omega=0$ which happens at the
$v=\sqrt{4\varepsilon^2+\gamma^2}$ point. If the coupling strength
is increased beyond this point, the peak of $\Lambda(\omega)$
becomes de-tuned from the peak of the temperature function and the
decoherence rate goes down. Since $\Lambda(\omega)$ plays the role
of a spectral function, the positions of its maxima
$\varepsilon_{\pm}$ give us the energies of charge states of the
system. Pushing them away from the Fermi level leads to  freezing
them out  so that the impurity is no longer a fluctuator and thus
does not contribute to the decoherence rate. Note that such a
strong coupling is still not a good news for the qubit operation
as the qubit and the impurity form together a four-level system
which, although remains coherent, does not operate as intended.

The above described behaviour of the decoherence rate as a
function of coupling strength is depicted in Fig.\ref{fig:gdep}.
Note that the maximum of $\Gamma_2$  has a cusp which is not
smeared by temperature (but would be smeared by including the
$\sigma_x$ part into the Hamiltonian):
\begin{figure}
\centering
\includegraphics[height=4cm]{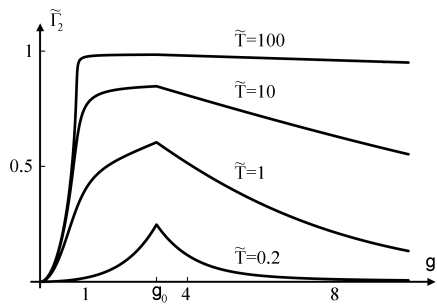}
\caption{Non-monotonic dependence of the decoherence rate on the
coupling strength at different temperatures for
$\tilde{\varepsilon}=3$; here $g_0=\sqrt{\tilde{\varepsilon}^2+1}$.}
\label{fig:gdep}
\end{figure}
Only at a rater high temperature ($\tilde{T}\sim 100$) the
decoherence rate practically saturates at its classical limit
$\Gamma_2^{{\text{(high)}}}=\gamma/2$.

\section{Relaxation in the FBC model}

Apart from decoherence, coupling of a qubit to the environment
also leads to relaxation processes, i.e.\ to a decay of diagonal
elements of the density matrix (\ref{rdm}), as outlined in Section
\ref{sec:dec}.  In the model under consideration,
Eq.~(\ref{themodel}), the presence of the $\hat{\sigma}_x$-part
inevitably induces relaxation. Such a relaxation would also
destroy coherence but the relaxation rate $\Gamma_1$ is typically
smaller than the decoherence rate $\Gamma_2$ and therefore the
time available for quantum operations is going to be determined by
the latter. Experiments on the relaxation rate are of special
interest as they can provide valuable information about the
degrees of freedom in the environment to which qubit is coupled.

It is convenient to diagonalize the qubit part of the Hamiltonian
(\ref{themodel}), transforming simultaneously the interaction part
$\hat{\sigma}_z$ into a sum of $\hat{\tau}_z$ and $\hat{\tau}_x$
terms - the Pauli matrices in the new basis:
\begin{align}
&\hat{H}=\frac{\Delta
E}{2}\,\hat{\tau}_z+\left(\cos\Theta\,\hat{\tau}_z+\sin\Theta\,
\hat{\tau}_x\right)\hat{V}+\hat{H}_{\text B}\,;\quad\,\,
\text{where}\nonumber\\
&\Delta E=\sqrt{(\delta E_c)^2+E_J^2}\,,\quad
\cos\Theta=\frac{\delta E_c}{\Delta E}\,,\quad
\sin\Theta=\frac{E_J}{\Delta E}\,.
\end{align}
As it is well known \cite{breuer}, in Born-Markov rotating-wave
approximation the decoherence rate $\Gamma_1$ is given by:
\begin{align}
\Gamma_1=\sin^2\Theta\,S(\Delta E)\,,\quad\quad
S(\Omega)=\left\langle\left\{\hat{V}(t)-
\langle\hat{V}\rangle,\hat{V}(t')-
\langle\hat{V}\rangle\right\}\right\rangle_{\Omega}\,,
\end{align}
where $S(\Omega)$ is the spectral density of noise. (The
appropriate derivation is outlined in Section \ref{sec:dec}). The
symbol $\langle\ldots\rangle_{\Omega}$ denotes the Fourier
transform of the bath-averaged expression with respect to $t-t'$
at the frequency $\Omega$.

\begin{figure}
\centering
\includegraphics[height=4cm]{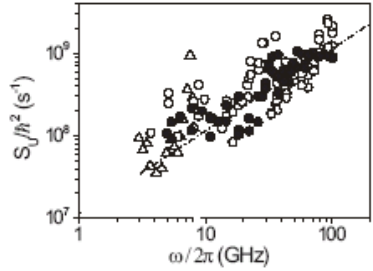}
\caption{The experimental data  by Astafiev et al \cite{astafiev}:
the noise spectral density  $S$ deduced from measurements of
$\Gamma_1$; open and closed symbols correspond to different
samples while circles and triangles -- to different measurement
regimes; $T\!\approx \!50 $mK.} \label{fig:astafiev}
\end{figure}

In the experiment of Astafiev et al \cite{astafiev} the authors
determined the spectral density of noise in a wide window of
frequencies, by changing $\Delta E$ and measuring the relaxation
rate. The experimental data for $S(\Omega)$ is presented in
Fig.\ref{fig:astafiev}.
Although the data points are somewhat scattered, the authors
nevertheless claim that they observed a linear behaviour of
spectral density of noise with peaks at $7$ and $23$GHz.

Below we will demonstrate that qualitatively the same behaviour is
expected from $S(\Omega)$ in the FBC model at low temperatures. A
contribution to the spectral density of noise from one impurity is
given by
\begin{align}
S(\Omega)&=\frac{v^2}{4}\,\Big\langle\left\{\hat{d}^\dagger
(t)\hat{d}(t)- \langle\hat{d}^\dagger
\hat{d}\rangle\,,\,\hat{d}^\dagger (t')\hat{d}(t')-
\langle\hat{d}^\dagger \hat{d}\rangle\right\}\Big\rangle_{\Omega}=\nonumber\\
&=\frac{\pi
v^2}{2}\,\coth\frac{\Omega}{2T}\int\limits_{-\infty}^{+\infty}
\dd\omega\,\nu_{\varepsilon}(\omega)\nu_{\varepsilon}(\omega+\Omega)
\left[n_F(\omega)-n_F(\omega+\Omega)\right]\,,\label{spectral-integral}
\end{align}
where the density of states $\nu_{\varepsilon}(\omega)$ is given by
(\ref{lorentzian}), and $n_F(\omega)$ is the Fermi distribution
function.

The frequencies accessed in the experiment \cite{astafiev} are in
the window $T\ll\Omega<100T$. Getting back to the rough estimates
for the probability of having an effective impurity with energy
smaller than temperature (the end of Section \ref{sec:fbcmodel}),
we see that even a scenario with $|\varepsilon_i|\gg 100T$  for
all effective impurities is quite likely. If this is the case then
the condition $T\ll\Omega\ll|\varepsilon|$ holds and it is clear
that the spectral density of noise $S(\Omega)$ exhibits linear
behaviour at such frequencies. As the thermal factor  is
practically equal to $1$ within the range $-\Omega<\omega<0$ and
zero otherwise, the smearing of this step is of the order of
temperature. Far to the right (assuming $\varepsilon$ to be
positive) we have two    peaks in DoS at $\varepsilon-\Omega$ and
$\varepsilon$. The main contribution to the integral
(\ref{spectral-integral}) is coming from the tails of
$\nu_{\varepsilon}$  at the interval $-\Omega<\omega<0$, rather
than from the thermal function tail in the vicinity of DoS peaks
which is exponentially suppressed. The latter contribution would
only win if $\gamma$ itself was exponentially small, but the
corresponding impurity is obviously non-effective. If $\gamma$ is
not too small,  we find
\begin{align}
S(\Omega)=\frac{\pi v^2}{2}\,\nu^2_{\varepsilon}(0)\Omega\quad\quad
T\ll\Omega\ll\varepsilon\label{linear-s}\,.
\end{align}
The summation of  contributions (\ref{linear-s}) coming from all
the effective impurities produces a linear in $\Omega$ function,
which can be a possible explanation for a linear (but very noisy)
trend in $S(\Omega)$ observed in \cite{astafiev}.

\begin{figure}
\centering
\includegraphics[height=4cm]{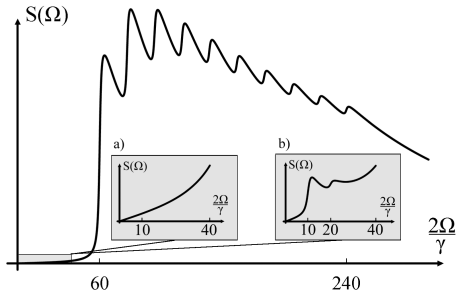}
\caption{The noise spectral density  $S(\Omega)$ in arbitrary
units at $T=0$ for the case of ten impurities with energies
uniformly distributed between $\tilde{\varepsilon}=60$ and
$\tilde{\varepsilon}=240$. The inserts   zoom  the bottom left
corner (the grey area on the main figure for
$\tilde{\varepsilon}\le 40$) when there is either a) no impurity
in this interval   or b) two impurities
($\tilde{\varepsilon}_1=10$, $\tilde{\varepsilon}=20$) with
coupling constants $v_{1,2}$ 50 times weaker than those for $60\le
\tilde{\varepsilon}\le 240$.} \label{fig:many-fluct}
\end{figure}
Humps in Fig.\ref{fig:astafiev} can be coming from less effective
but more numerous impurities which makes the situation
$T\ll|\varepsilon|<100T$ possible. At the point
$\Omega\sim\varepsilon$ the left one of the DoS double-Lorentzian
overlaps with the thermal function thus producing a peak in
$S(\Omega)$. As $\Omega$  further increases, the overlap between
the left Lorentzian and the thermal function stays essentially the
same but the region where it occurs drifts away to the left from
the second Lorentzian which is stationary positioned at
$\omega=\varepsilon$. Qualitatively the picture of a linear
behaviour of $S(\Omega)$ followed by a peak at
$\Omega\sim\varepsilon$ is the same for any relation between
$\gamma$ and $T$, but the form of the peak is different.

With a purely illustrative purpose  we show in
Fig.\ref{fig:many-fluct} a picture corresponding to the following
scenario: all the tunneling rates are the same, $\gamma_i=\gamma$,
obeying $\gamma\gg T$ (which effectively allows one to put $T$ to
zero), and several impurities are uniformly distributed in some
interval of energies. All these impurities contribute to the
linear behaviour on the left of this interval [the grey area
zoomed in the insert a)], while humps might be due to fluctuators
with smaller energies weakly coupled to the qubit could not break
the general linear trend [the grey area zoomed in the insert b)].
The behaviour of $S(\Omega)$ in the grey area qualitatively
resembles the experimental data.

\section{Conclusions}

In these lectures, we have described some essential features of
loss of coherence by a qubit coupled to the environment. We have
first presented well known semiclassical arguments that relate
both decoherence and relaxation to the environmental noise. Then
we have shown that models with pure decoherence (when there is no
relaxation in qubit states as that part of coupling to the
environment that leads to flipping the states is excluded) can be
exactly solvable. As an example, we have treated in detail the
model of fluctuating background charges
[4--7] 
 which is believed to
describe one of the most important channels for decoherence for
the charge Josephson junction qubit. Following our earlier
treatment \cite{gyl2}, we have shown that the decoherence rate
saturates at `high' temperatures \cite{paladino1,galperin1} while
becoming linear in $T$ at low temperatures and showing in all
regimes a non-monotonic behaviour as a function of the coupling of
the qubit to the fluctuating background charges. We have also
considered, albeit only perturbatively, the qubit relaxation by
the background charges and demonstrated that the quasi-linear
behaviour of the spectral density of noise deduced from the
measurements of the relaxation rate can be qualitatively explained
within this model in the low temperature regime.

We thank B.~L.~Altshuler, Y.~M. Galperin, R.~Fazio and  A.
Shnirman for useful comments.   This work was supported by the
EPSRC grant GR/R95432.

\bibliographystyle{myprsty}

\end{document}